# Atomic resolution interface structure and vertical current injection in highly uniform MoS$_2$ heterojunctions with bulk GaN

F. Giannazzo[1]*, S. E. Panasci[1], E. Schilirò[1], G. Greco[1], F. Roccaforte[1], G. Sfuncia[1], G. Nicotra[1], M. Cannas[2], S. Agnello[2,1,3], E. Frayssinet[4], Y. Cordier[4], A. Michon[4], A. Koos[5], B. Pécz[5]

[1] Consiglio Nazionale delle Ricerche – Istituto per la Microelettronica e Microsistemi (CNR-IMM), Z.I. VIII Strada 5, 95121, Catania, Italy
e-mail: filippo.giannazzo@imm.cnr.it

[2] Department of Physics and Chemistry Emilio Segré, University of Palermo, Via Archirafi 36, 90143 Palermo, Italy

[3] AtenCenter, University of Palermo, Viale delle Scienze Ed.18, 90128 Palermo, Italy

[4] Université Côte d'Azur, CNRS, CRHEA, 06560, Valbonne, France

[5] Centre for Energy Research, Institute of Technical Physics and Materials Science, Budapest, Hungary

## Abstract

The integration of two-dimensional MoS$_2$ with GaN recently attracted significant interest for future electronic/optoelectronic applications. However, the reported studies have been mainly carried out using heteroepitaxial GaN templates on sapphire substrates, whereas the growth of MoS$_2$ on low-dislocation-density bulk GaN can be strategic for the realization of "truly" vertical devices. In this paper, we report the growth of ultrathin MoS$_2$ films, mostly composed by single-layers (1L), onto homoepitaxial n$^-$-GaN on n$^+$ bulk substrates by sulfurization of a pre-deposited MoO$_x$ film. Highly uniform and conformal coverage of the GaN surface was demonstrated by atomic force microscopy, while very low tensile strain (~0.05%) and a significant p$^+$-type doping (~4.5×10$^{12}$ cm$^{-2}$) of 1L-MoS$_2$ was evaluated by Raman mapping. Atomic resolution structural and compositional analyses by aberration-corrected electron microscopy revealed a nearly-ideal van der Waals interface between MoS$_2$ and the Ga-terminated GaN crystal, where only the topmost Ga atoms are affected by oxidation. Furthermore, the relevant lattice parameters of the MoS$_2$/GaN heterojunction, such as the van der Waals gap, were measured with high precision. Finally, the vertical current injection across this 2D/3D heterojunction has been investigated by nanoscale current-voltage analyses performed by conductive atomic force microscopy, showing a rectifying behavior with an average turn-on voltage

$V_{on}$=1.7 V under forward bias, consistent with the expected band alignment at the interface between $p^+$ doped 1L-MoS$_2$ and n-GaN.

## 1. Introduction

In the last years, the integration of two dimensional (2D) transition metal dichalcogenides, such as the molybdenum disulfide (2H-MoS$_2$) [1], with wide-bandgap (WBG) semiconductors, including silicon carbide (4H-SiC) [2,3], gallium nitride (GaN) [4,5] and AlGaN alloys [6], has been the object of increasing research interest. In fact, the combination of the unique physical properties of ultrathin MoS$_2$ films, such as the tunable energy bandgap [7,8] and the relatively high in-plane carrier mobility [9], with the robust electronic properties of SiC and GaN (wide-bandgap, high breakdown field and electron saturation velocity) [10,11] paves the way to novel device concepts, including heterojunction diodes based on the vertical current injection at the 2D/3D semiconductors interfaces [12,13], and optoelectronic devices (e.g., dual wavelength photodetectors) exploiting the optical response of these materials in the visible and UV spectral ranges [14,15]. Due to this wide application potential, several approaches have been explored to fabricate MoS$_2$ heterojunctions with SiC and GaN, including exfoliation of MoS$_2$ flakes from bulk crystals [6,14,15], transfer of MoS$_2$ thin films grown on foreign substrates [12,13], and the direct MoS$_2$ growth on SiC and GaN substrates by scalable approaches like chemical vapor deposition (CVD) [16,17] and pulsed laser deposition (PLD) [18,19,20]. In particular, the epitaxial growth of MoS$_2$ on these hexagonal symmetry WBG substrates is favored by the very low lattice parameter mismatch in the basal planes, ~2.9% in the case of SiC(0001) [20] and <1% for GaN(0001) [16].

Recent reports on heterojunction diodes formed by monolayer (1L) or few layers MoS$_2$ grown on 4H-SiC bulk crystals already showed very promising performances for vertical devices applications [20, 21]. On the other hand, although MoS$_2$ integration with GaN has been more widely investigated, the majority of studies has been performed using heteroepitaxial GaN films grown on insulating sapphire substrates [16,17]. Despite the high density of threading dislocations (~$10^8$-$10^9$ cm$^{-2}$) [22], originating from the lattice and thermal expansion coefficient mismatch with the substrate, low-cost heteroepitaxial GaN on sapphire is widely employed in optoelectronics, whereas GaN grown on Si and on semi-insulating SiC are employed in medium-voltage (650 – 900 V) power transistors and RF transistors, respectively. On the other hand, low-dislocation-density (<$10^5$ cm$^{-2}$) n-GaN homoepitaxial layers on n$^+$ bulk GaN [22] are necessary for the implementation of vertical transistors and diodes, allowing to fully exploit the potential of GaN at higher voltage/current rating [23]. Although free-standing GaN still remain high-cost substrates, large improvement in the bulk growth methods have been made during last decade [24], and high quality bulk GaN wafers up to 100 mm diameter are now available.

As a matter of fact, the high dislocation density in heteroepitaxial GaN on sapphire is expected to significantly affect the rectification properties of the MoS$_2$/GaN interface. In this respect, the integration of ultra-thin MoS$_2$ films on low defectivity homoepitaxial GaN layers can be extremely beneficial, both to explore the "intrinsic" rectification properties of the MoS$_2$/GaN interface and for the future development of 2D/3D vertical devices for high power and high frequency applications. However, only few studies have been reported so far about MoS$_2$ growth on bulk GaN, and they have been limited to the case of multilayer MoS$_2$ on a n$^+$ doped ($10^{18}$ cm$^{-3}$) GaN substrate [25].

In this paper we report on the highly uniform growth of ultrathin MoS$_2$, mostly formed by monolayer (1L) or bilayers (2L), on top of homoepitaxial GaN by a two-steps chemical vapor deposition (CVD) approach, consisting in the sulfurization of a pre-deposited ultrathin MoO$_x$ film. As compared to the commonly employed single-step CVD with S and MoO$_3$ vapours, typically resulting in the nucleation and growth of isolated triangular domains of MoS$_2$, this method allows to obtain continuous MoS$_2$ films on large area with good thickness uniformity by controlling the initial MoO$_x$ thickness [21,26]. In particular, the surface morphology of as-grown MoS$_2$ films was demonstrated to be very conformal to the atomic steps of n-GaN homoepitaxy. Atomic resolution structural and compositional analyses, carried out by aberration-corrected transmission electron microscopy combined with electron energy loss spectroscopy (EELS), revealed a nearly-ideal van der Waals (vdW) interface between $MoS_2$ and the Ga-terminated GaN crystal, where only the topmost Ga atoms are affected by oxidation. Furthermore, these analyses provided a precise evaluation of the relevant lattice parameters of the heterojunction, such as a vdW gap ranging from 4.71 to 5.06 Å. These atomic scale investigations were complemented by micro-scale Raman mapping with high statistics, which revealed a very low tensile strain at $MoS_2$/GaN interface, consistently with the lattice matching of the two crystals, and a significant $p^+$-type doping of $MoS_2$, ascribed to $MoO_x$ residues. Finally, the vertical current injection across this 2D/3D heterojunction has been investigated by local current-voltage analyses performed by conductive atomic force microscopy (C-AFM), which showed a rectifying behavior with an average turn-on voltage $V_{on}$=1.7 V under forward bias, consistent with the expected band alignment at $p^+$ $MoS_2$/n-GaN interface.

## 2. Experimental details

The starting material for our $MoS_2$ growth experiments was a ~300 μm thick GaN substrate, Ga polarity (0001) oriented and $n^+$-doped ($N_D$-$N_A$=1-5×$10^{18}$ $cm^{-3}$), covered by a ~3 μm thick $n^-$ GaN homoepitaxial layer ($N_D$-$N_A$≈1×$10^{16}$ $cm^{-3}$). An heteroepitaxial GaN template, grown by metal organic chemical vapor deposition (MOCVD) on sapphire, was also employed for benchmarking purposes. A $MoO_x$ film (with ~1.5 nm thickness, evaluated by AFM-step height measurements) was deposited on the GaN surface by DC magnetron sputtering from a Mo target using a Quorum Q300-TD equipment, followed by natural oxidation in air. Afterwards, the sulfurization process was carried out in a two-heating horizontal furnace, hosting a crucible with S powders in the low temperature zone ($T_1$=150 °C) and the $MoO_x$/GaN sample in the high temperature zone ($T_2$=700°C). Argon was used as the carrier gas to transport the S vapour to the sample's surface and the process duration was 1 hour. The same sulfurization conditions have been recently employed for $MoO_x$ films with similar thickness on 4H-SiC, resulting in the formation of a highly uniform 1L-2L $MoS_2$ film [21].

The composition of the as-deposited $MoO_x$ films (with x=2.8) before sulfurization was preliminary evaluated by XPS analyses, carried out using an XSAM 800 instrument by Kratos Analytical, with a non-monochromatic Mg Kα X-ray source (energy = 1253.6 eV). Mo 3d core level spectra are reported in Fig.S1 of the Supplementary Information, from which a stoichiometry $MoO_x$, with x=2.8, was deduced. The coverage uniformity of the $MoS_2$ films on GaN has been first investigated by tapping mode AFM with a DI-3100 system by Bruker using ultra-sharp Si tips (~5 nm curvature radius). Furthermore, the distribution of the number of $MoS_2$ layers, strain and doping were evaluated by micro-Raman spectroscopy and mapping with a 532 nm laser wavelength, carried out using a WiTec Alpha and a Horiba equipment. The structural and chemical properties of the $MoS_2$ films on GaN have been investigated at atomic resolution by transmission electron microscopy on cross-sectioned lamellas, prepared by focused-ion-beam (FIB). First, high-resolution transmission electron microscopy (HRTEM), scanning transmission electron microscopy (STEM) in high-angle-annular-dark-field (HAADF) mode, and energy dispersion spectroscopy (EDS) analyses were acquired using an image corrected ThermoFisher THEMIS 200 microscope with 200 keV electron beam. Furthermore, atomic resolution analyses of the $MoS_2$/GaN interface by STEM in the HAADF and annular bright field (ABF) modes, and chemical mapping by electron energy loss spectroscopy (EELS) have been performed using a probe corrected JEOL ARM 200F microscope at primary electron beam energy of 200 keV. Finally, the vertical current injection in $MoS_2$ heterojunctions with bulk GaN was investigated by nanoscale resolution current-voltage (I-V) analyses in the front-to-back configuration performed by conductive AFM (C-AFM) with a DI-3100 system by Bruker, using Pt-coated conductive tips.

3. **Results and discussion**

Fig.1(a) shows a representative AFM morphological image of the as- grown $MoS_2$ on the n⁻ homoepitaxial GaN sample, confirming the formation of a homogenous film, highly conformal to the atomic steps of the GaN surface, as evident from the higher resolution image in Fig.1(b). The scanned area in Fig.1(a) includes some scratches, intentionally performed in the film in order to evaluate its thickness by step height measurements. Fig.1(c) shows a height line profile, extracted along the red-line in Fig1(a), from which ~0.7 nm film thickness is deduced in the considered region, which is in the range of typical thickness values for 1L-MoS$_2$ measured by AFM. This value is also in good agreement with the distance between the topmost S plane of MoS$_2$ and the topmost Ga plane of the (0001) oriented GaN, evaluated by atomic resolution STEM in this paper.

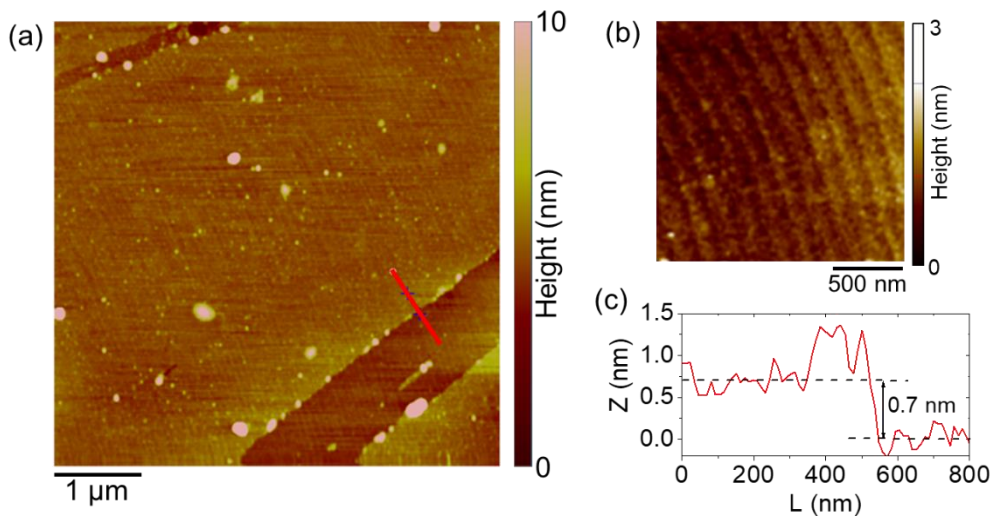

Figure 1.- (a) AFM image showing the as-grown MoS$_2$ on GaN morphology. The scanned area includes scratches intentionally performed in the MoS$_2$ film to evaluate its thickness (b) Higher resolution image showing conformal coverage on the GaN surface atomic steps. (c) Line profile extracted along the red line indicated in the panel (a).

In the following, the results of Raman mapping will be reported, to evaluate the distribution of the number of MoS$_2$ layers, the doping and strain on large areas and with high statistics. A typical Raman spectrum of the MoS$_2$ heterostructure with bulk GaN is shown in Fig.2(a), where the vibrational peaks of MoS$_2$ ($E_{2g}$ and $A_{1g}$) and GaN ($E_2$ high energy and $A_1$(LO)) are indicated by red and blue boxes, respectively. Furthermore, Fig.2(b) reports a detail of the MoS$_2$ spectrum, showing the out-of-plane ($A_{1g}$) and the in-plane ($E_{2g}$) modes located at 406.0±0.6 cm$^{-1}$ and 384.6±0.6 cm$^{-1}$, respectively. In addition to the two main $E_{2g}$ and $A_{1g}$ modes, the presence of additional spectral contributions, i.e. the LO(M) mode at ~380.5 cm$^{-1}$ and ZO(M) mode at ~414.5 cm$^{-1}$, has been revealed by the deconvolution analysis in Fig.2(b). These contributions, typically associated to $MoS_2$ lattice defects [27], can be related to the grain boundaries of the polycrystalline $MoS_2$ layers produced by sulfurization of $MoO_x$ films deposited by evaporation or sputtering. The typical size of $MoS_2$ crystalline domains in these layers, evaluated by plan-view TEM analyses, is in the range from few tens of nm to ~100 nm, depending on the substrate and on the sulfurization temperature [28, 29, 30, 31, 32]. Similarly, we estimated a domain size of 50 – 60 nm for our $MoS_2$ films transferred on a TEM grid.

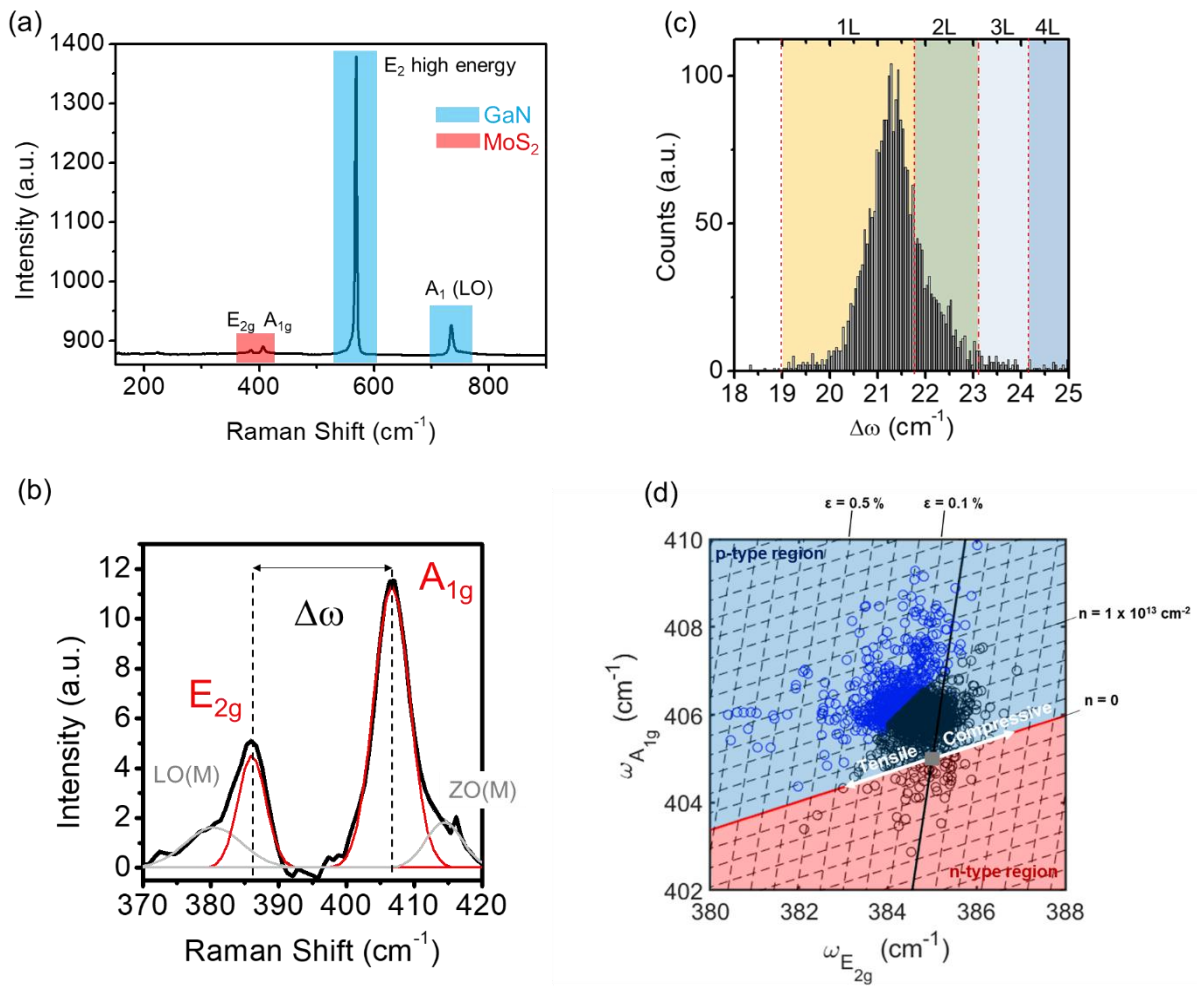

Figure 2.- (a) Raman spectrum of the 2D/3D heterojunction with the corresponding $MoS_2$ (red square) and GaN (blue squares) peaks. (b) Detail on the $MoS_2$ in-plane ($E_{2g}$) and out-of-plane ($A_{1g}$) peaks after the deconvolution analysis from which the presence of further peaks defined LO(M) and ZO(M) can be appreciated. (c) Distribution of the difference ($\Delta\omega = \omega_{A1g} - \omega_{E2g}$) between the two $MoS_2$ Raman peaks obtained by maps on 2500 arrays of spectra over a 10 μm×10 μm sample area. The number of $MoS_2$ layers corresponding to the different $\Delta\omega$ ranges is indicated in the upper horizontal scale. (d) Correlative $E_{2g}$ vs $A_{1g}$ plot to extract information on the strain and doping characterizing the $MoS_2$, due to the interaction with GaN

bulk substrate. Black points correspond to 1L-MoS$_2$ regions, whereas the blue points correspond to thicker MoS$_2$ regions.

The wavenumber difference $\Delta\omega=\omega_{A_{1g}}-\omega_{E_{2g}}$ between the A$_{1g}$ and E$_{2g}$ peaks is commonly used to estimate the number of MoS$_2$ layers [33,34]. Hence, to evaluate the MoS$_2$ thickness uniformity on the bulk GaN, we performed a statistical analysis over an array of 50×50 Raman spectra collected on a 10 μm×10 μm sample area. The result of this analysis is represented by the $\Delta\omega$ histogram in Fig.2(c). As indicated in the upper horizontal scale, the $\Delta\omega$ values in the range between ~19 and ~21.7 cm$^{-1}$ correspond to 1L-MoS$_2$, the values between ~21.7 and ~23.1 cm$^{-1}$ to 2L-MoS$_2$, the values between ~23.1 and ~24.2 cm$^{-1}$ to 3L-MoS$_2$, and the values and >24.2 cm$^{-1}$ to 4L-MoS$_2$ and thicker films. By integrating the counts in each $\Delta\omega$ range, a 1L-MoS$_2$ percentage of ~73%, 2L-MoS$_2$ percentage of ~24%, 3L-MoS$_2$ percentage of ~2%, and 4L-MoS$_2$ percentage of ~1% have been estimated.

In addition to the determination of the number of layers, the positions of the two main Raman peaks allow to evaluate the strain and doping of the ultra-thin MoS$_2$ membrane [35], resulting by the growth conditions and the interaction with the GaN substrate. Fig.2(d) shows a correlative plot of the A$_{1g}$ vs E$_{2g}$ peak positions extracted from the array of Raman spectra. In particular, the ($\omega_{A_{1g}}$, $\omega_{E_{2g}}$) data points belonging to 1L-MoS$_2$ (i.e. those with $\Delta\omega<21.7$ cm$^{-1}$) have been indicated by black dots, and the data points belonging to thicker (2L, 3L, 4L) MoS$_2$ regions have been indicated by blue dots. The red and black lines are the strain and doping lines, respectively, which represent the theoretical correlations between the vibrational frequencies of a purely strained and of a purely doped 1L-MoS$_2$ [36,37,38]. These two lines cross in a point (grey square) corresponding to the E$_{2g}$ and A$_{1g}$ wavenumbers for an ideally unstrained and undoped 1L-MoS$_2$ [37]. Here, the literature values $\omega_{E_{2g}}=385$ cm$^{-1}$ [37] and $\omega_{A_{1g}}=405$ cm$^{-1}$ [37] for a suspended 1L-MoS$_2$ membrane have been assumed as the best approximation for this ideal case. The experimental points are centered in the upper (blue) region of the plot, indicating a significant p-type doping of MoS$_2$. In particular, from the black data points an average hole density p≈4.5×10$^{12}$ cm$^{-2}$ was estimated for 1L-MoS$_2$. Furthermore, 1L-MoS$_2$ exhibits a very small tensile strain (average value ~0.05%), consistently with the small mismatch (<1%) between the in-plane lattice parameters of MoS$_2$ and GaN [2]. The origin of the p-type doping can be ascribed to the presence of MoO$_x$ residues in the MoS$_2$ thin films produced by the sulfurization process. This aspect has been explained and demonstrated by XPS analysis in our previous work, in which we employed the same growth conditions for the MoS$_2$ on 4H-SiC [21].

It is worth mentioning that, for benchmarking purposes, similar micro-Raman mapping and statistical analyses have been carried out on MoS$_2$ films grown under identical conditions on the GaN-on-sapphire template. The results of this investigation (reported in Fig.S2 of the supporting information) indicated very similar distributions in the MoS$_2$ layers number, doping and strain, demonstrating the growth process reproducibility on homo- and hetero-epitaxial GaN samples.

After this statistical characterization of the as-grown MoS$_2$ film on micrometer scale areas, the structural and chemical properties of MoS$_2$ interface with homoepitaxial GaN were investigated at nanometric/atomic scale by transmission electron microscopy and spectroscopy. Fig.3(a) shows a low magnification cross-sectional TEM image, confirming the presence of a continuous layered film, composed by 1L or 2L MoS$_2$, on the GaN surface. The amorphous carbon film with the nanocrystalline Pt on top served as protection for MoS$_2$ layers during FIB preparation. Noteworthy, TEM analyses at the same magnification, carried out on MoS$_2$ films grown on the GaN-on-sapphire templates (see Fig.S3 in the Supporting Information), showed the same MoS$_2$ film coverage, consistently with Raman results, while the main difference consisted in the presence of threading dislocations in the heteroepitaxial GaN crystal.

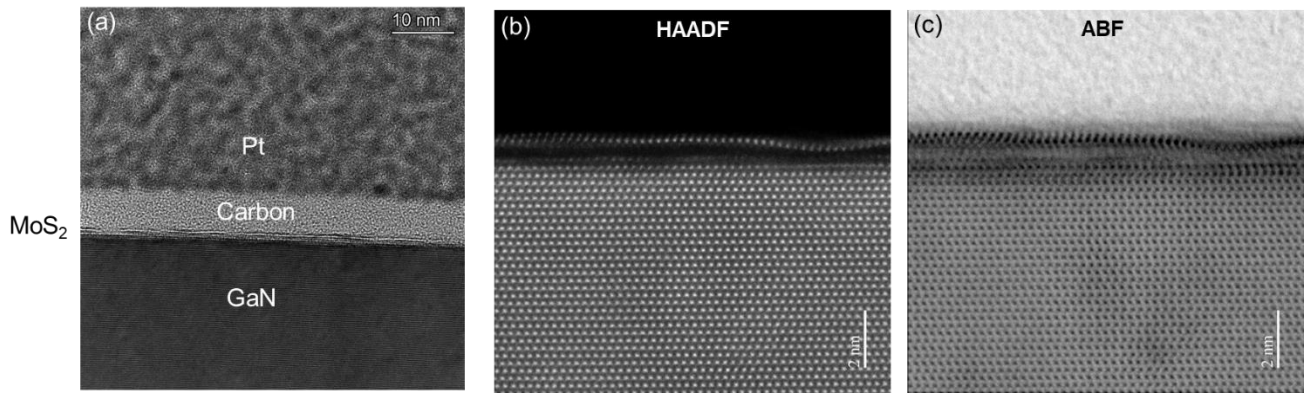

Figure 3.- (a) Low magnification TEM image of the MoS$_2$ film grown onto the n$^-$ GaN homoepitaxial layer on the bulk GaN substrate. Atomic resolution STEM images acquired in the HAADF mode (b) and ABF mode (c)

To get atomically resolved information on the MoS$_2$/GaN interface structure, STEM analyses have been simultaneously performed in the HAADF and the ABF imaging modes, as reported in Fig.3(b) and (c), respectively. In the HAADF image, obtained by collecting electrons scattered at large angles, the bright contrast is associated to the columns of atoms (i.e., Mo, Ga) with high atomic number (Z), whereas the low Z atoms (i.e., N and S) are difficult to be imaged. On the other hand, in the ABF imaging mode both atoms with high and lower Z can be simultaneously visualized as dark spots with higher and lower intensities, respectively. The combination of these two complementary imaging modes allowed a precise evaluation of the relevant lattice distances in the heterostructure, as shown later in this paper. Looking in detail to the near-interface GaN lattice, a first interesting feature that can be observed in the HAADF images is a reduced intensity of the topmost Ga atomic plane with respect to the underlying GaN crystal. Previously reported STEM investigations of MoS$_2$ on GaN (grown by single-step CVD at 800 °C) also reported on the presence of a "modified" GaN surface region, formed by two GaN planes underneath MoS$_2$ [17]. Such modification was tentatively explained by the authors in terms of the GaN surface reconstruction or its partial oxidation during the MoS$_2$ growth, but these hypotheses were not conclusively supported by the reported electron microscopy analyses, due to the insufficient resolution and lack of chemical information [17].

Here, the compositional properties of the $MoS_2$/GaN heterostructure were investigated in detail by EELS and elemental mapping (as shown in Fig.4) and by EDS (as reported in Fig.S4 of the Supporting Information). Fig.4(a) displays an overview EELS spectrum, collected in the region shown in the HAADF spectrum image of Fig.4(b). The spectral features associated to the main atomic species present in the heterostructure, i.e, the Ga-M and Ga-L edges, the N-K edge, the S-L edge and the Mo-M edge, are clearly detected. Furthermore, the presence of a weak intensity O-K peak was also observed. The chemical maps, showing the spatial distribution of the different atomic species in the spectrum image region, are reported in Fig.4(c)-(h). Besides confirming $MoS_2$ formation, these analyses demonstrated that the near-surface GaN region is almost unaffected by the sulfurization of the pre-deposited $MoO_x$ film at 700 °C, except for the incorporation of oxygen traces. The presence of a very small amount of oxygen also in the $MoS_2$ region can be deduced from the EELS map of Fig.4(g), and it is confirmed by the EDS analyses in Fig.S3. It can be associated to some $MoO_x$ residues still present after the sulfurization, as previously observed by XPS analyses on samples prepared under similar conditions [21]. Such small percentages of $MoO_x$ can be relevant for the electronic properties of $MoS_2$, since they can produce a p-type doping of the layer, as revealed by Raman analyses (see Fig.2(d)).

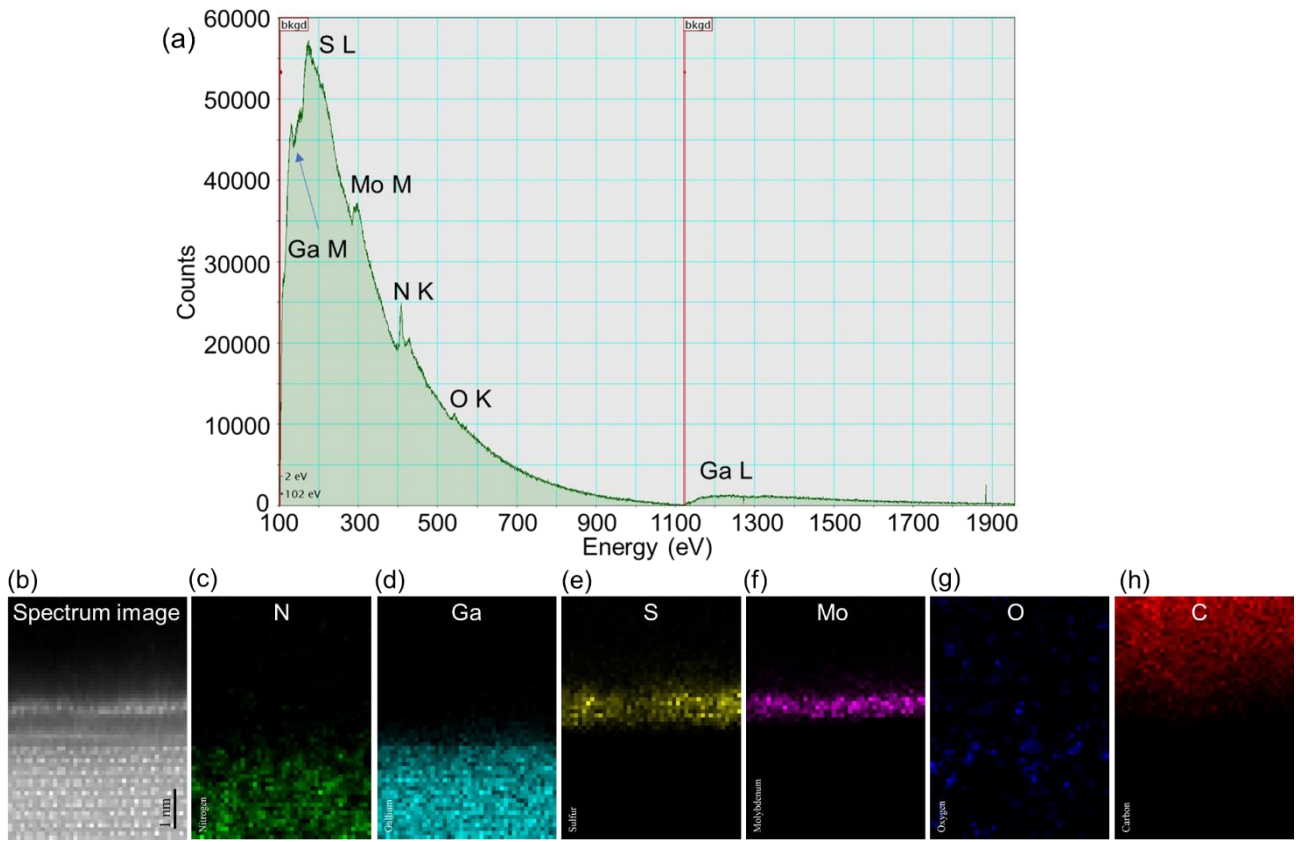

Figure 4.- (a) EELS spectrum of the MoS$_2$/GaN heterostructure, with the indication of edges of the detected atomic species. HAADF spectrum image of the analyzed region (b) and corresponding elemental maps of Nitrogen (c), Gallium (d), Sulfur (e), Molybdenum (f) Oxygen (g) and Carbon (h).

To get further information on the atomic structure of near interface region, Fig.5(a) and (c) show two atomic resolution STEM analyses, simultaneously collected in the HAADF (a) and ABF (c) imaging modes, respectively. The intensity line profiles, averaged on the individual atomic planes, are superimposed on the left and right sides of the two images. Accurate estimations of the lattice distances in the direction perpendicular to the interface were obtained by measuring the distances between the peaks in these line profiles.

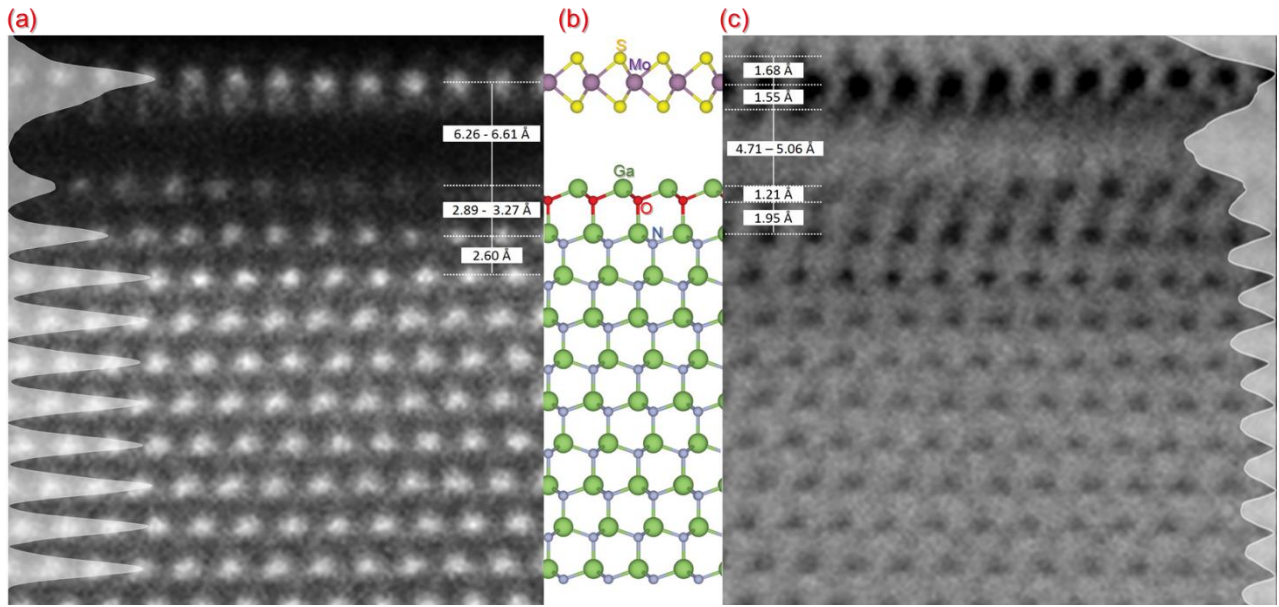

Figure 5.- Atomic resolution cross-sectional STEM of 1L-MoS$_2$/GaN heterostructure in HAADF (a) and ABF (c) imaging modes. The intensity line profiles, averaged on the individual atomic planes, are superimposed on the left and right sides of the two images. (c) Schematic representation of the atomic arrangement in the heterostructure. The Mo-Ga and Ga-Ga inter-plane distances were evaluated from the HAADF intensity profiles. The S-Mo-S distances within 1L-MoS$_2$, the S-Ga van der Waals gap, and the Ga-O distance in GaOx surface were evaluated from the ABF intensity profiles.

Firstly, the peak in the HAADF line profile corresponding to the topmost Ga plane exhibits a significantly lower intensity as compared to the underlying Ga atomic planes. Furthermore, the separation between the 1$^{st}$ and the 2$^{nd}$ Ga planes (ranging from 2.89 to 3.27 Å) is significantly larger than the one between the 2$^{nd}$ and the 3$^{rd}$ Ga planes (~2.60 Å), which is the same for the underlying bulk GaN crystal. A distance $d_{Ga-Mo}$ between the topmost Ga plane of GaN and the plane of Mo atoms of the MoS$_2$ film ranging from 6.26 to 6.61 Å was also evaluated from the HAADF intensity line-profile in Fig.5(a). Further insights on the S-Mo-S distances within 1L-MoS$_2$, and the S-Ga distance between the bottom S atoms of MoS$_2$ and the topmost Ga plane of the substrate have been deduced from the ABF image and intensity line profile in Fig.5(c). Firstly, we observed slightly different distances ($d_{S-Mo}$=1.68 Å and $d_{Mo-S}$=1.55 Å) between the central Mo atoms and the upper and lower S planes of 1L-MoS$_2$. Furthermore, a vdW gap $d_{S-Ga}$ ranging from 4.71 to 5.06 Å between the lower S plane and the topmost Ga was evaluated from Fig.5(c). Finally, the larger separation between the 1$^{st}$ and $2^{nd}$ Ga atomic plane was explained by the substitution of oxygen to nitrogen atoms, i.e. by the formation of $GaO_x$, as schematically illustrated in Fig.5(b). This view is also consistent with the results of chemical analysis reported in Fig.4.

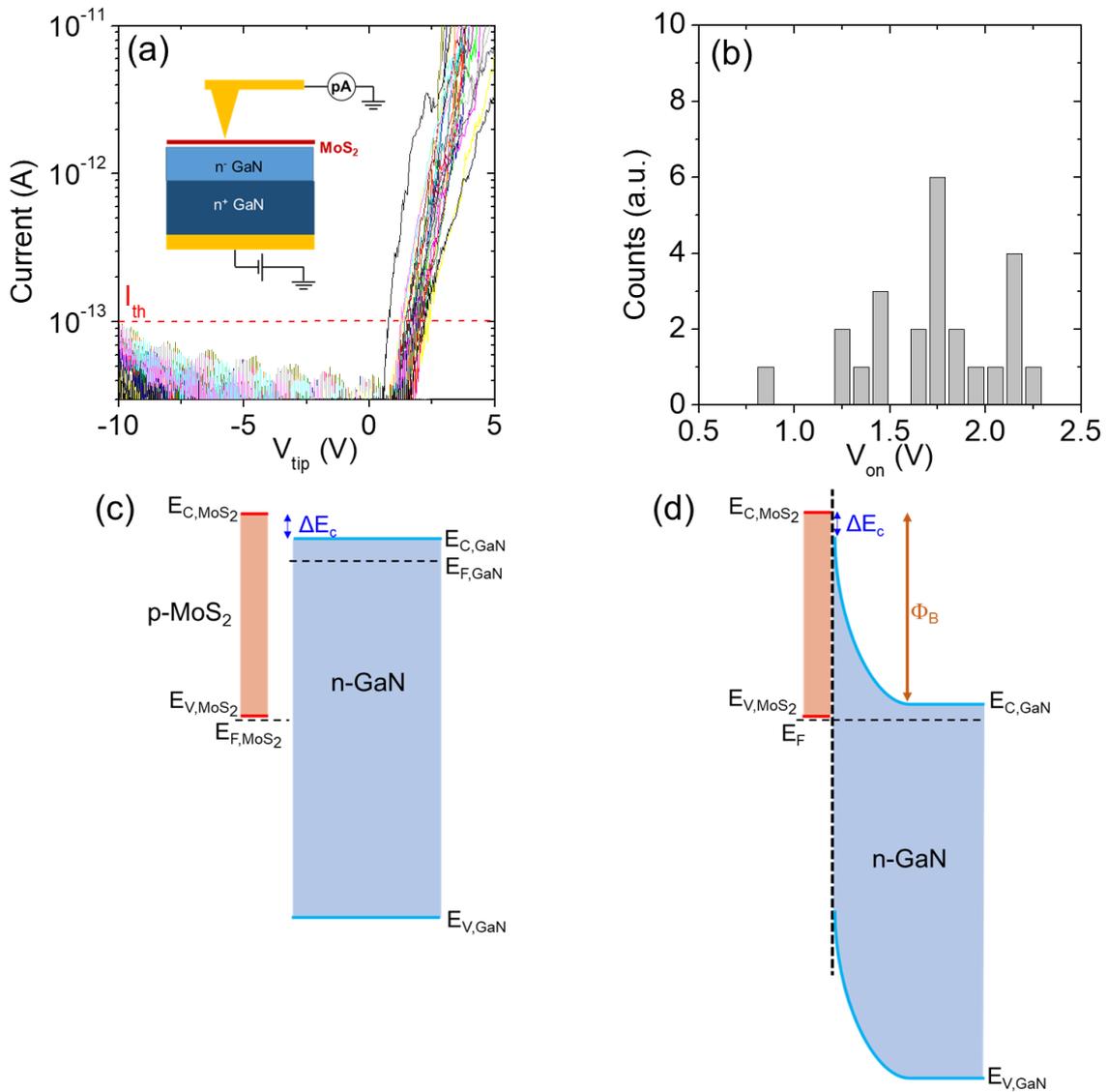

Figure 6.- (a) Set of current-voltage (I-$V_{tip}$) characteristics collected by C-AFM in the front-to-back configuration (see insert) over an array of 25 tip positions on the $MoS_2$/GaN heterojunction. (b) Distribution of the onset voltage extracted from the I-$V_{tip}$ curves at a fixed current threshold of $10^{-13}$ A. Energy band diagrams of the heterojunction before (c) and after (d) contact formation under equilibrium conditions.

After assessing the strain, doping and interface structural properties of ultrathin MoS$_2$ on bulk GaN, the electrical properties of MoS$_2$ films and the vertical current injection across the MoS$_2$/GaN vdW heterojunction were investigated by nanoscale resolution current-voltage analysis based on C-AFM. The current injection from the C-AFM tip to MoS$_2$ was preliminary evaluated, as discussed in Fig.S5 of the Supporting Information, showing a behavior consistent with p$^+$-doping of MoS$_2$. Afterwards, front-to-back current measurements from the nanometric Pt tip contact with MoS$_2$ surface to the macroscopic Ni/Au back-contact with n$^+$- bulk GaN, have been performed, according to the configuration illustrated in the insert of Fig.6(a). A dc bias ramp is applied between these two contacts, while the flowing current is measured by a high sensitivity current amplifier connected to the tip. Fig.6(a) shows a set of local I-V$_{tip}$ curves measured on an array of 5×5 tip positions on 5μm×5μm area. As can be seen, the I-V$_{tip}$ curves, reported on a semi-log scale, showed a rectifying behavior, with a negligible current under reverse bias and an exponential rise of the current under forward bias above an onset voltage (V$_{on}$). The histogram in Fig.6(b) represents the distribution of the V$_{on}$ values extracted from the I-V$_{tip}$ curves array at a fixed value of the threshold current I$_{th}$=1×10$^{-13}$ A (above the noise level), as indicated in Fig.6(a). This distribution exhibits an average value of ~1.7 V, with a standard deviation of ±0.4 V. The very low reverse current and the relatively high onset voltage in the I-V$_{tip}$ curves indicates a high energy barrier for current injection at MoS$_2$/GaN interface.

Fig.6(c) schematically illustrates the energy band diagram of the heterojunction before the 1L-MoS$_2$/GaN contact formation. The type II band alignment is deduced from recent literature reports based on XPS/UPS analyses of the 1L-MoS$_2$/GaN system, where conduction band discontinuity ΔE$_c$ values in the range from 0.07 to 0.6 eV have been evaluated [4,39]. The Fermi level position (E$_F$-E$_{C,GaN}$≈-0.11 eV) in the n$^-$ GaN epitaxy was estimated from the nominal doping level (N$_D$≈1×10$^{16}$ cm$^{-3}$) according to the relation E$_F$-E$_{C,GaN}$=kT/q ln(N$_D$/N$_C$), being N$_C$=1.8×10$^{18}$ cm$^{-3}$ the effective density of states in the conduction band at T=300 K for GaN [40], k the Boltzmann constant and q the electron charge. On the other hand, the average hole density (p≈4.5×10$^{12}$ cm$^{-2}$) in the p-type doped

1L-MoS$_2$, evaluated by Raman mapping in Fig.2(d), corresponds to an average concentration of N$_h$≈1.4×10$^{20}$ cm$^{-3}$, assuming that this charge is confined within an effective thickness of ~3.23 Å for 1L-MoS$_2$, as evaluated by the ABF TEM (see Fig.5(c)). Such a value is larger than the effective density of states in the valence band N$_v$≈1.2×10$^{19}$ cm$^{-3}$ for 1L MoS$_2$, estimated by using a parabolic valence-band model and the corresponding analytical formula N$_V$=2(2πm$_h$kT/h$^2$)$^{3/2}$, where h is the Planck constant, T=300 K, and m$_h$=0.61m$_e$ is the holes effective mass for 1L-MoS$_2$. Hence, a degenerately p-type-doped 1L-MoS$_2$, with the Fermi level very close to the valence band edge (or inside the valence band) can be assumed, as schematically depicted in Fig.6(c). Finally, the energy band diagram configuration of this p$^+$/n$^-$ 2D/3D heterojunction after contact formation (under equilibrium conditions) is schematically illustrated in Fig.6(d), where Φ$_B$ indicates the energy barrier after the Fermi levels alignment. According to this band diagram, Φ$_B$ can be estimated as Φ$_B$=E$_{g,1L-MoS2}$-|E$_F$-E$_{C,GaN}$|, being E$_{g,1L-MoS2}$ the bandgap of 1L-MoS$_2$ (i.e., 1.8-1.9 eV) and |E$_F$-E$_{C,GaN}$|≈0.11 eV. This value is consistent with the average onset voltage (V$_{on}$=1.7 V) previously extracted from the I-V$_{tip}$ characterization, while the spread (± 0.4 V) of the local V$_{on}$ values (histogram in Fig.6(d)) can be ascribed to local variations of MoS$_2$ p-type doping, resulting in local changes of the energy band alignment.

4. Conclusion

In conclusion, the growth of uniform MoS$_2$ films, mostly composed by single-layers, onto homoepitaxial n$^-$-GaN on n$^+$ bulk substrates has been obtained by sulfurization at 700 °C of a pre-deposited MoO$_x$ film. The grown MoS$_2$ layers are highly conformal coverage to the GaN surface atomic steps. They exhibit very low tensile strain (~0.05%), consistently with the low in-plane lattice mismatch between MoS$_2$ and GaN, and a significant p$^+$-doping (average holes density p≈4.5×10$^{12}$ cm$^{-2}$), that was ascribed to residual MoO$_x$ from the sulfurization process. A nearly-ideal vdW interface between MoS$_2$ and the Ga-terminated GaN crystal, where only the topmost Ga atoms are affected by oxidation, was demonstrated by atomic resolution STEM in the HAADF and ABF modes, combined with EELS. The relevant lattice parameters of the MoS$_2$/GaN heterojunction were measured with high precision. Finally, the vertical current injection across this 2D/3D heterojunction was investigated by nanoscale current-voltage analyses performed by conductive AFM, showing a rectifying behavior with an average turn-on voltage V$_{on}$≈1.7 V under forward bias, consistent with the expected band alignment at p$^+$-MoS$_2$/n-GaN interface.

The obtained results on the integration of highly uniform and ultrathin p$^+$ MoS$_2$ films with low-dislocation density n$^-$ GaN homoepitaxial layers, and the demonstration of the rectifying electrical behavior of this 2D/2D heterostructure are important steps towards the development of vertical heterojunction diodes for high-power and high-frequency applications.

**Aknowledgements**

S. Di Franco (CNR-IMM, Catania) is acknowledged for the expert technical assistance with sample preparation. We thank N. Nemeth (EK) for XPS measurements, and K. Kandrai, P. Kun, G. Zoltan Radnoczi (EK) for their contribution in MoS$_2$ films transfer experiments and TEM analyses. P. Fiorenza (CNR-IMM, Catania), F. M. Gelardi (Univ. of Palermo), M. Al Khalfioui (Université Côte d'Azur, CNRS-CRHEA) are acknowledged for useful discussions.

This work has been funded, in part, by MUR in the framework of the FlagERA-JTC 2019 project ETMOS. Funding from CNR/HAS bilateral project GHOST-II for traveling is also acknowledged. E.S. acknowledges funding from the ECSEL JU project GaN4AP (Grant Agreement No. 101007310). B.P. and A.K. acknowledge funding from the national projects TKP2021-NKTA-05 and NKFIH K_134258. The CNR-IMM and Univ. of Palermo acknowledge funding from the "SiciliAn Micro and NanO TecHnology Research and InnovAtion Center (Samothrace)". Part of the experiments was carried out using the facilities of the Italian Infrastructure Beyond Nano.